# Effect of homogenization on precipitation behavior and strengthening of 17-4PH stainless steel fabricated using laser powder bed fusion

Kun Li [a,1], Soumya Sridar [a], Susheng Tan [b,c], Wei Xiong [a,*]

[a] Physical Metallurgy and Materials Design Laboratory, Department of Mechanical Engineering and Materials Science, University of Pittsburgh, Pittsburgh, PA 15261, USA
[b] Department of Electrical and Computer Engineering, University of Pittsburgh, Pittsburgh, PA 15260, USA
[c] Petersen Institute of Nanoscience and Engineering, University of Pittsburgh, Pittsburgh, PA 15260, USA
[1] Present Address: School of Mechanical Engineering, Chongqing University, China
* Corresponding Author, Emails: w-xiong@outlook.com and weixiong@pitt.edu
Tel: +1 (412) 383-8092, Fax: +1 (412) 624-4846

**Abstract**

Effective post-heat treatment is critical to achieve desired microstructure for high-performance in additively manufactured (AM) components. In this work, the influence of homogenization on microstructure-property relationship in 17-4PH steels has been investigated. Precipitation of NbC, oxides, and ε-Cu were observed in the as-built 17-4PH steels. To design an optimum post-heat treatment, homogenization was performed at 1050°C for different times followed by aging at 482°C for 1 hour. It was identified that homogenization for 1 hour followed by aging leads to the best combination of strength and ductility due to the refinement of martensite and prior austenite grains. Improved tensile properties were achieved for the post-heat-treated alloys that exceeded the traditionally fabricated 17-4PH steels. Through comprehensive microstructure characterization, it was deduced that the incoherent ε-Cu precipitates in the as-built alloy were dissolved through homogenization, and subsequently, re-precipitated as coherent Cu-rich clusters during aging. This study demonstrates that altering the precipitation behavior using post-heat treatment is an effective pathway to significantly improve the mechanical properties of AM alloys.

# 1. Introduction

17-4 precipitation-hardening (PH) stainless steel possesses high strength owing to precipitation strengthening by Cu-rich clusters in the martensitic matrix as well as excellent corrosion resistance [1-3]. Hence, this alloy has been employed in aerospace and marine applications as well as in power plants and chemical industries involving thermal exposure up to 315°C [4-8]. The wrought 17-4PH steel usually has a fully martensitic structure with a minor amount of δ-ferrite [9,10]. The traditional heat treatment for 17-4PH steel consists of homogenization which can eliminate the δ-ferrite followed by aging that can precipitate coherent Cu-rich clusters leading to strengthening and transform to incoherent ε-Cu precipitates with FCC structure due to over-aging [1,11-14]. The fabrication and post-heat treatment of 17-4PH steels has gained more interest in order to achieve better tradeoff between the microstructure and mechanical properties [15-17].

In comparison with the traditional manufacturing methods where a component is fabricated by material removal, additive manufacturing (AM) is a state-of-the-art technique based on layer-by-layer deposition of a material [16-19]. Laser powder bed fusion (LPBF) is a powder-based AM technique to print 3-dimensional parts directly from the metal powder with advantages such as complex near-net shaping and absence of extra tooling [20,21]. Due to these advantages, this method has been applied for fabricating 17-4PH steels [22-24]. However, the complex interactions during the LPBF process will result in defects such as cracks, porosity, and poor surface finish, if the process is not well controlled [25-27]. Besides, the anisotropic heat conduction during the cyclic heating and cooling generates a heterogeneous microstructure leading to anisotropic properties [28-31].

The reports available in literature pertaining to additive manufacturing of 17-4PH stainless steel using LPBF technique mainly focuses on three different aspects: (a) effect of processing parameters [32-36], (b) effect of powder atomizing methods [37-40] and (c) post heat treatments [41-44]. The effect of processing parameters, for instance, energy density, hatch spacing and scanning strategy on the characteristics of microstructure and properties were widely investigated [24,32-34]. The results showed that it had a close correlation between energy density/input, porosity and mechanical properties along with the microstructure [32-34]. Gu *et al*. [32] proposed that low laser power and scanning speed were beneficial for the LPBF processing of 17-4PH steels. The layer thickness has a significant effect on porosity, while the hatch spacing was closely related to the scanning speed [24]. It was also reported that double scan strategy can be utilized to achieve high density and hardness for the as-printed samples than single scan [35]. The scanning direction can lead to the change in porosity and fraction of retained austenite during the melting process [36]. In relation to the influence of atomizing media on 17-4PH steels fabricated by LPBF process, it has been reported that water atomized powder had superior surface finish and uniformity in deposition [37-39], while it increased carbon content along with the formation of a dual-phase microstructure consisting of martensite and austenite [8]. On the other hand, different gas atomizing atmospheres could lead to different fraction of martensite and austenite [5,6,40]. In comparison with argon atomized powder, nitrogen atomized powder led to higher austenite fraction since nitrogen is an austenite stabilizer [40].

Due to the inhomogeneity and anisotropy introduced by laser melting, post-heat treatment is inevitable for LPBF processed 17-4PH steels. As summarized in Fig. 1 [9,15,41,42], two-step heat treatment is applied for wrought 17-4PH stainless steels. Typically, the homogenization temperature is required to be above 1000$^o$C to provide a fully supersaturated martensite. The aging temperature and time are usually altered to balance strength and toughness according to specific applications [7,15,40]. Aging heat treatment performed at 482$^o$C for 1 h improves the hardness and strength, while aging at 593$^o$C for 4 h improves the toughness and ductility [2]. The precipitation sequence and microstructure evolution during aging has been studied extensively for traditionally manufactured 17-4PH steels [3,11,43]. However, there are only few works that investigate the effect of homogenization on the precipitation behavior during aging [7,40,44]. In particular, additively manufactured 17-4PH steels are expected to have a distinct difference in recrystallization and precipitation behavior due to the anisotropic microstructure developed during the processing. The evolution of microstructure during homogenization is critical for the following aging heat treatment which has not been studied till now for 17-4PH steels fabricated using LPBF process.

In this work, different homogenization heat treatments followed by an identical aging treatment was performed with the help of thermodynamic simulations, to explore the effect of homogenization on 17-4PH steel manufactured using LPBF technique. The recrystallization and precipitation kinetics of additively manufactured 17-4PH steels before and after post-heat treatment were thoroughly investigated with the help of extensive high-resolution microstructural characterization. The mechanical properties were determined to investigate the influence of precipitation behavior during homogenization on the strength. The underlying correlation between precipitation during homogenization and in the following aging step has been elucidated.

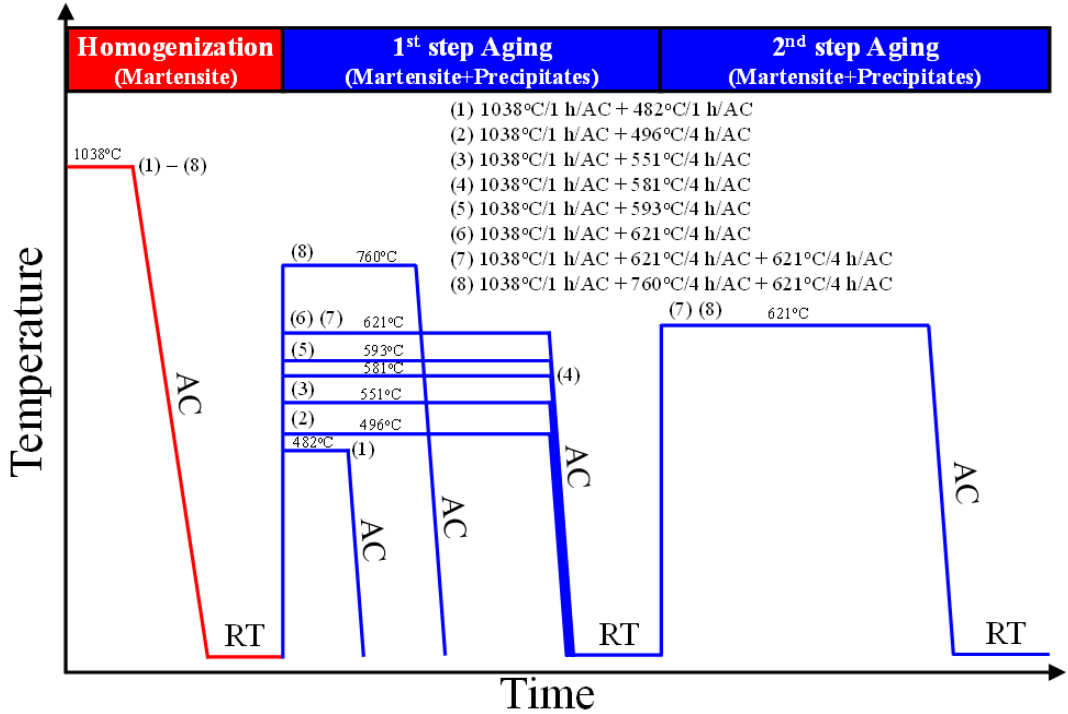

Fig. 1. Schematic time-temperature map showing the different post-heat treatments performed for wrought 17-4PH stainless steel [9,15,41,42].

## 2. Experiments and modeling

### *2.1 CALPHAD-aided post-heat treatment design*

Based on the powder composition listed in Table 1, the equilibrium phase fraction as a function of temperature was calculated using the Thermo-Calc software with commercial multicomponent database for steel (TCFE v9) as shown in Fig. 2. From this plot, it is clearly evident that above 970°C, the major secondary phases get dissolved and those phases that are beneficial for pinning the grain boundaries during homogenization such as carbide are present. Hence, a homogenization temperature of 1050°C was chosen such that it is high enough to dissolve all the major secondary phases while, it is not too high to promote δ-ferrite formation.

Table 1. The nominal composition (in wt.%) of 17-4PH stainless steel powder.

| Cr | Ni | Cu | Nb | Mn | Si | C | O | N | Fe |
|---|---|---|---|---|---|---|---|---|---|
| 15.84 | 4.55 | 3.87 | 0.37 | 0.32 | 0.36 | 0.019 | 0.05 | 0.01 | Bal. |

For the subsequent aging process, the size, distribution and coherency with the matrix for Cu clusters are the critical factors influencing mechanical properties and corrosion resistance [1,45]. According to Fig. 2, it is obvious that Cu clusters form from 950°C and reaches the stable volume fraction of 2.5% at 450°C. An aging temperature between 450 and 650°C is considered to have the strengthening effect [46-48] and in addition, other phases such as σ, G and $M_{23}C_6$ may form in this range, which influences the fatigue and corrosion resistance [49,50]. As the major focus of the present work is on the effect of homogenization, an identical aging treatment was performed at 480°C for 1 h after different homogenization treatments since, it is reported to be the peak aging condition for 17-4PH steels [2,46].

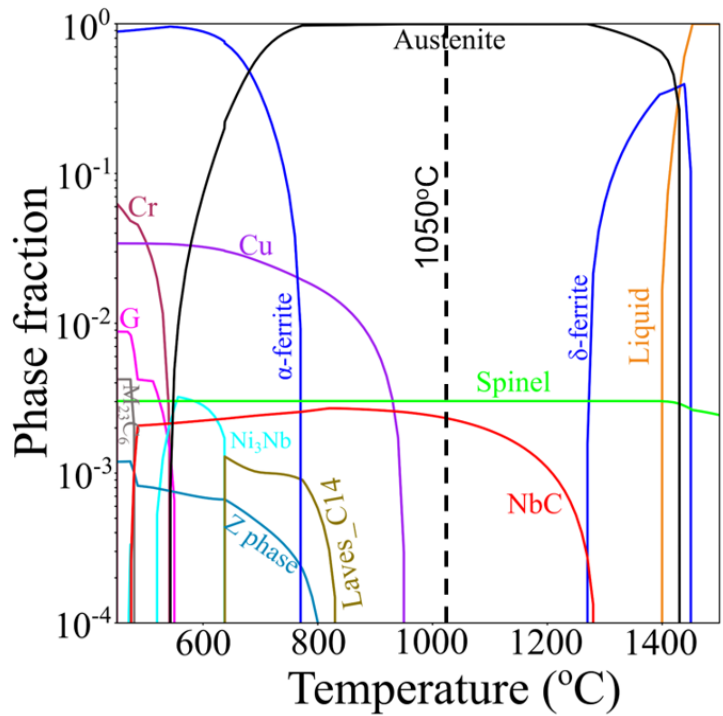

Fig. 2. Calculated equilibrium phase fraction as a function of temperature for 17-4 PH steel powder composition.

## 2.2 Experiments

The argon atomized pre-alloyed 17-4PH steel powder manufactured by Kennametal Co. in the size range of 45 – 53 μm were printed using EOS M 290 Machine (Electro-Optical-Systems company, Germany) in an argon atmosphere. The printing parameters were as follows: Laser power = 195 W; Scanning speed = 900 mm/s; Layer thickness = 0.04 mm; Hatch spacing = 0.1 mm. Several cubes (10×10×10 mm³) and tensile bars designed according to ASTM standard E8 were printed using these parameters. The samples were encapsulated inside quartz tubes under vacuum and back-filled with pure argon for heat treatment. Subsequently, the samples were homogenized at 1050°C for different times (0.5, 1, 2, 3, 4, 6, 8 and 10 hours), followed by air cooling and further aging at 482°C for 1 h followed by air cooling.

The samples were mechanically polished, and the as-polished surface was etched with Swede's reagent (5 mL HCl + 25 mL HNO$_3$ + 6 g FeCl$_3$ + 50mL distilled water). Optical microscopy (OM, Zeiss Axio Lab A1) and scanning electron microscopy (SEM, Zeiss Sigma 500 VP) combined with energy dispersive X-ray spectroscopy (EDS, Oxford Aztec) were employed to observe the microstructure and determine the compositions. Electron backscattered diffraction (EBSD, FEI Scios DualBeam FIB SEM equipped with EDAX Hickory EBSD system) scans were carried out to investigate the phase evolution during different homogenization heat treatments and were analyzed using TSL-OIM Analysis™ v8 software package. The reconstruction of prior austenite grains (PAG) was performed by the ARPGE software [51,52] with the EBSD data as input. The Kurdjomov and Sachs (K-S) orientation relationship [53] was used due to the maximum reconstruction percentage. The linear intercept method was used to measure the reconstructed PAG size. High resolution microstructure characterization was performed using transmission electron microscope (TEM, JEOL JEM-2100F TEM) operated at 200 kV using thin foils electropolished in a solution of 10% perchloric acid and 90% ethanol with a voltage of 30 V at −30°C. Hardness was evaluated using LECO LM800 Vickers microhardness tester. The load was set as 300 grams with a dwell time of 10 seconds. The reported values are an average of 10 measurements. Tensile tests were performed using MTS 880 universal testing machine with 100 kN capacity at room temperature with a crosshead speed of 1 mm/min. SEM imaging in secondary electron mode combined with EDS was performed to observe the fracture surface after the tensile tests.

## 3. Results

### *3.1 Microstructure of as-built 17-4PH alloy*

Figure 3 shows the microstructure of as-built 17-4PH steel observed using OM and the orientation maps obtained from EBSD (X – Normal direction, Y – Transverse (scan) direction and Z – Build direction). The scanning patterns in XY and XZ planes are clearly visible, with a hatch spacing of 100 μm (Fig. 3a2) and layer thickness of 40 μm (Fig. 3a4). These values are consistent with the printing parameters indicating that the printing process was well controlled. The melt pools overlapped with each other, which formed an extremely regular structure consisting of big bright grains centered in the melt pools and small gray blocks surrounding them, as shown in Figs. 3a3 and 3a5 while, there are very few submicron circular pores (Fig. 3a5).

The porosity of printed samples was measured to be ~0.5% using the Archimedes method demonstrating that highly dense 17-4PH steels were successfully printed using LPBF in the present study. However, the as-built sample had very strong texture as exhibited by the EBSD orientation maps in different planes (Figs. 3b1-3b5). In the XZ plane, the grains grew exactly parallel to the build direction since it is the preferred direction for high thermal dissipation during the printing (Fig. 3b4). Numerous small and equiaxed sub-grains formed surrounding the large columnar grains. These sub-grains were constrained along the boundaries of melt pools in XY plane (Fig. 3b3). This results from the effect of several thermal cycles and rapid cooling on the nucleation and recrystallization of grains in the overlapped areas of different layers.

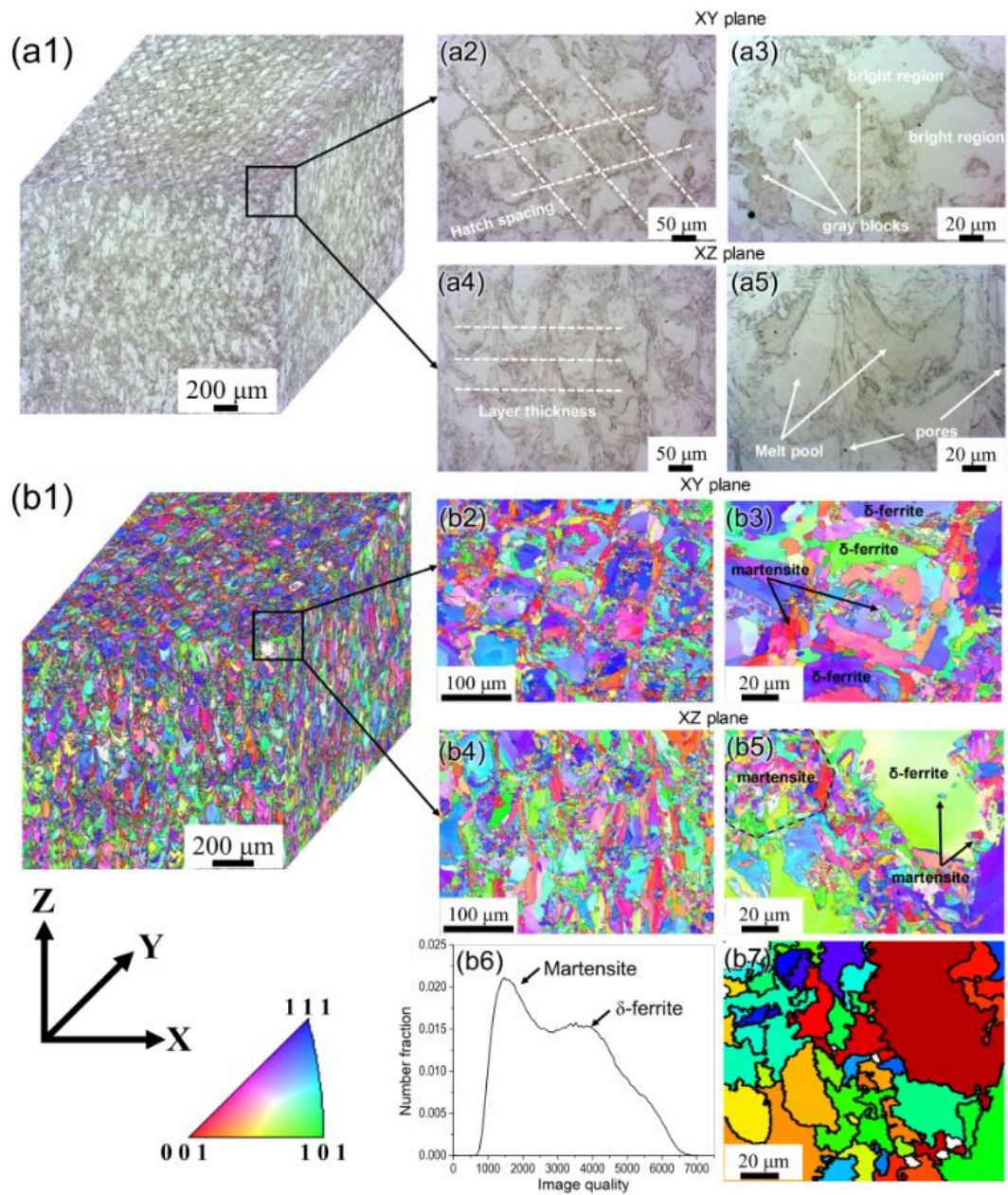

Fig. 3. Microstructure of as-built 17-4PH steel: OM analysis: (a1) 3D OM microstructure reconstruction, (a2 and a3) OM micrographs in XY plane, (a4 and a5) OM micrographs in XZ plane; EBSD analysis: (b1) 3D EBSD microstructure reconstruction, (b2 and b3) EBSD orientation maps in XY plane, (b4 and b5) EBSD orientation maps in XZ plane, (b6) Image quality plot of (b5) showing the presence of both martensite and δ-ferrite, and (b7) Reconstructed PAG map for the as-built 17-4PH steel.

To identify the matrix phase in the as-built alloy, the phase fraction was analyzed using EBSD (Fig. 3b5). The as-built alloy has a BCC structure with an average grain size of 20±1.7 μm. The image quality (IQ) plot of the BCC phase shows two peaks (Fig. 3b6), which indicates two kinds of BCC structures are present in the as-built alloy. The peak with higher IQ value corresponds to δ-ferrite and the other one is martensite, because δ-ferrite has very lower dislocation density in comparison with martensite. The formation of δ-ferrite in as-built alloy is due to the extremely rapid cooling rate of melt pool during printing, which retains the δ-ferrite from high temperature to room temperature. It is located in the center of melt pools, as shown in Figs. 3b3 and 3b5, while martensite has formed in the overlapped areas of melt pools because of remelting from multiple thermal cycles. The volume fraction of δ-ferrite was calculated to be 36.7±4.3% using image analysis.

In addition to the matrix phase in the as-built alloy, the precipitates that formed during the LPBF process in 17-4PH steels were observed using SEM in backscatter electron (BSE) mode as shown in Fig. 4. There were two types of precipitates in the as-built alloy. The bright particles were seen as clusters with an average size of 40 nm, while the dark particles were dispersed in the matrix with uniform size of 30 nm (Fig. 4b). The EDS composition profiles for these two types of precipitates are shown in Figs. 4c and 4d. The bright particles correspond to Cu and the dark particles indicate the presence of oxide rich in Si. According to the thermodynamic calculation for 17-4PH steel powder (Fig. 2), it could be Spinel/Rhodonite oxides, which was also found in a recent study of stainless steel 316L fabricated by LPBF method [29]. The oxygen content of the powder, as indicated in Table 1, is high enough to promote the formation of oxide particles [29]. As the cooling rate is as high as $10^6$ K/s during the LPBF process, the Rhodonite particles can precipitate as fine round dispersoids [29].

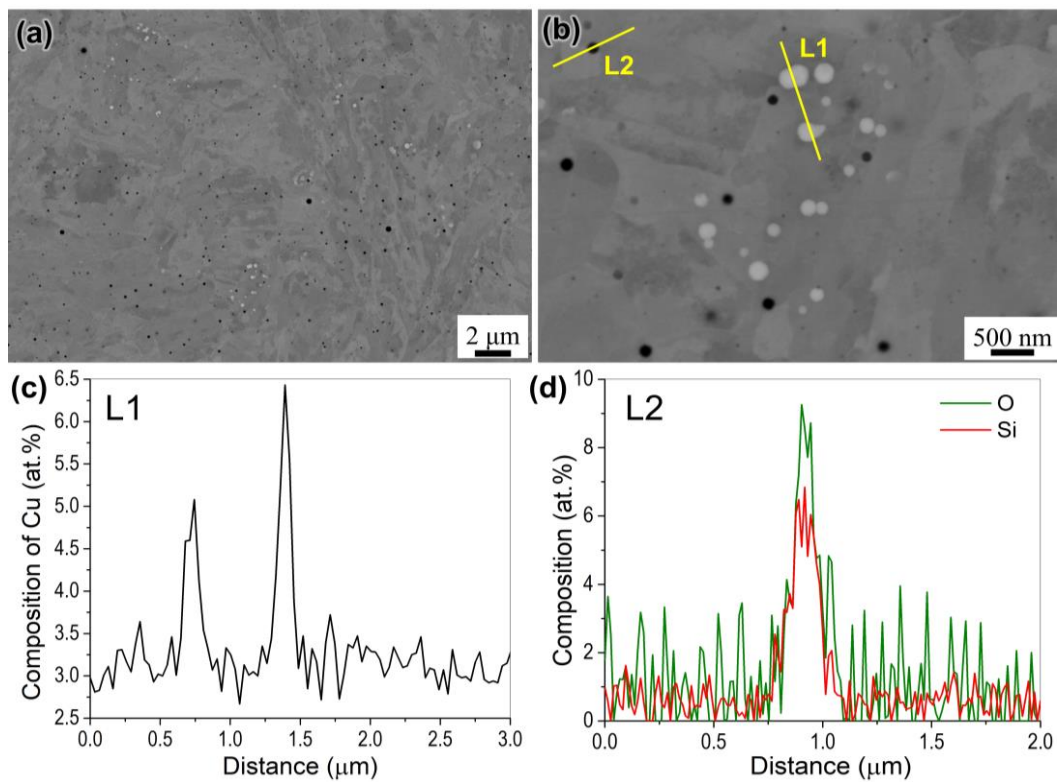

Fig. 4. Precipitates in as-built 17-4PH steel with EDS analysis: (a) BSE-SEM micrograph of precipitate distribution, (b) high magnification micrograph of (a), (c) EDS composition profile of bright particles along L1 and (d) EDS composition profile for dark particle along L2.

To further clarify the nature of precipitates in as-built 17-4PH steels, TEM was used to characterize the microstructure (Fig. 5). Fig. 5a shows the overall morphology of martensite in the as-built alloy. The width of martensite lath is ~2 μm, with parallel orientation and the laths were embedded into each other. Enormous amount of dislocations tangled within the martensite lath indicates a high stored energy/residual stress owing to the AM process. Dark particles (yellow dashed circles) surrounded by dislocation intertwining were observed at a higher magnification (Fig. 5b), which are oxides as identified by the selected area electron diffraction pattern (SAED). These particles were further analyzed using the scanning transmission electron microscopy (STEM), as shown in Fig. 5c. The precipitates were seen as bright and dark circles at high angle annular dark field (HAADF) mode that are rich in Mn, Si and O and poor of Fe and Ni, demonstrating that these are oxides. However, the bright particle is rich in Cr and Mn, while the dark particle shows lesser amount of Cr and Mn. This implies that there are oxides of two different chemistries, showing different contrast in the dark field mode. It has been reported that the two oxides that are typically present in additively manufactured steels are Rhodonite ($MnSiO_3$) which is a relatively unstable phase usually observed in Mn/Si deoxidized steels, and Spinel ($MnCr_2O_4$) which is more stable than the Rhodonite [54,55]. As shown in the STEM maps (Fig. 5c), the dark particles show lesser Cr and rich of Si while, the bright particles are rich in both Cr and Si. Based on the chemical composition of Rhodonite and Spinel, the dark particles correspond to Rhodonite phase, and the bright particles are a mixture of Rhodonite and Spinel. The transformation from unstable Rhodonite to stable Spinel is possibly due to the tempering effect during the LPBF process with cyclic heating/cooling below the solidus temperature. These oxides can act as the nuclei for dislocations and impede their migration as shown in Fig. 5b. At a higher magnification as shown in Figs. 5d and 5e, another nano-particle cluster was observed in the matrix (as indicated by blue dashed circle), which are Cu precipitates according to the EDS analysis. Furthermore, there was no coherency between the precipitate and matrix as seen from the high-resolution TEM micrograph (Fig. 5f) since it is ε-Cu particles with FCC structure. The straight dislocations got spiraled when reaching the Cu precipitate, which indicates that the dislocations were nucleated and emitted from these precipitates since they were not coherent with the matrix. Therefore, the pre-existing ε-Cu particles are incoherent and thus, they are not suitable for the strengthening of 17-4 steel.

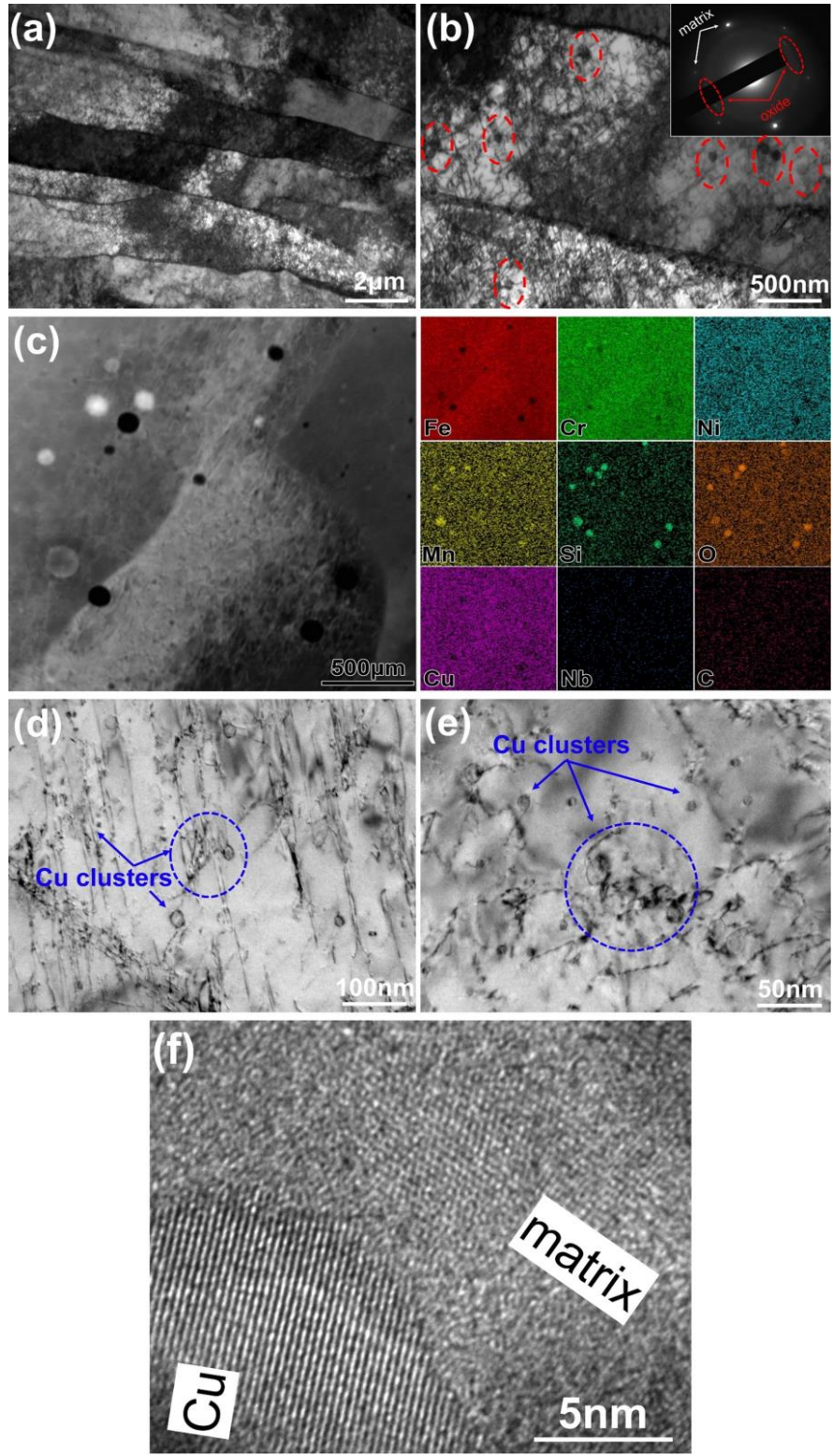

Fig. 5. Characterization of precipitates using TEM in as-built 17-4 PH steel: (a) parallelly oriented martensite, (b) high magnification micrograph of (a) showing dislocations and oxide particles (inset image showing SAD pattern), (c) HAADF image (left) and corresponding EDS maps for different elements (right), (d) Cu clusters and dislocations in the matrix, (e) high magnification micrograph of (d) and (f) high resolution image of Cu cluster in the matrix showing incoherency.

## 3.2 Homogenization studies for additively manufactured 17-4PH steels

Homogenization heat treatment was performed at 1050°C between 0.5 to 10 hours for as-built 17-4 PH steel since the incoherent ε-Cu precipitates need to be dissolved, δ-ferrite needs to be eliminated and residual stresses that cumulate during the LPBF processing needs to be annihilated. Figure 6 shows the microhardness of 17-4PH steels homogenized at 1050°C for different times. It is obvious that the as-built alloy had the highest hardness (328 HV). This is due to the presence of Cu precipitates and high density of dislocations in the martensite lath as observed from the high-resolution TEM micrographs (Fig. 5). The hardness decreased significantly after 1 hour of homogenization and further, increased and reached a peak at 4 hours. Subsequently, the hardness decreased rapidly till 10 hours. On the other hand, the PAG size shows an inverse behavior as it decreased till 1 hour of homogenization and further, increased rapidly till 10 hours. The points marked with red arrows are the representative homogenization durations (0, 0.5, 1, 4, and 8 h) that is associated with some critical variation of microstructural attributes during homogenization. The inverse pole figure (IPF) and reconstructed PAG maps for the homogenized 17-4PH steels are shown in Fig. 7. It can be observed that the alloys were comprised of fine lath martensite without obvious preferred orientations and the δ-ferrite was eliminated after homogenization. The coarsening of the prior austenite grains can be observed from the reconstructed PAG maps.

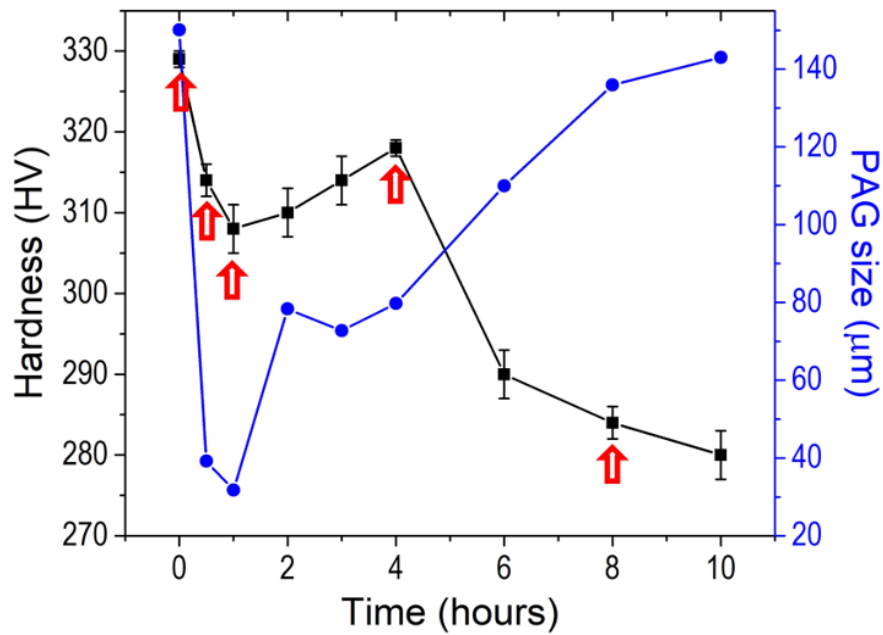

Fig. 6. Variation in hardness and PAG size as a function of homogenization time (*Red arrows represent the critical homogenization times that are analyzed further in the following discussion*)

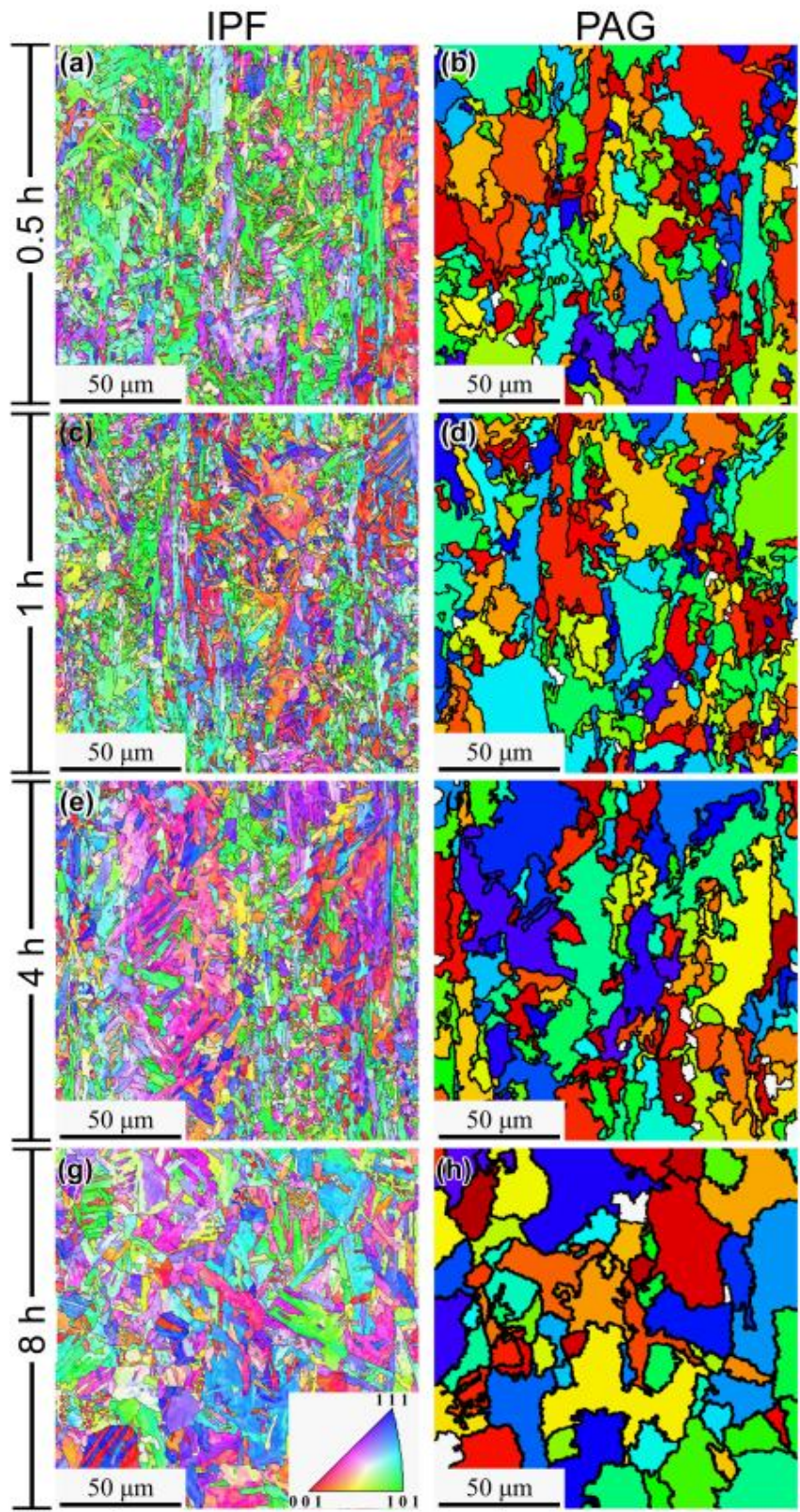

Fig. 7. IPF and reconstructed PAG maps for additively manufactured 17-4PH steels homogenized at 1050°C for (a, b) 0.5, (c, d) 1, (e, f) 4 and (g, h) 8 hours.

In addition to recrystallization, the precipitation behavior during homogenization was also studied by observing the microstructure in SEM, as shown in Fig. 8. According to the equilibrium phase fraction versus temperature plot (Fig. 2), there are two secondary precipitates namely, NbC and oxide at 1050°C. Correspondingly, in Fig. 8, the bright particles are NbC, and the dark particles are oxides. The EDS composition profiles show that the bright particles are rich in Nb and the dark particles are rich in Si and O. During initial stages of homogenization (0.5 and 1 hour) there was no obvious change in the oxide distribution in comparison with the as-built alloy. However, increased precipitation of NbC was observed, mostly distributed near the grain boundaries, while some fine and circular particles stayed in the matrix after 0.5 hours of homogenization (Fig. 8a). These particles could be the pre-existing Cu retained in the matrix due to the limited diffusion during the short homogenization time. After longer homogenization (4 and 8 hours), the oxides were coarsened with a rapid increase in the size of NbC and it got mixed with oxides, as presented in Figs. 8c and 8d. Some NbC particles precipitated on the site of the oxides and enveloped them (see composition profiles of L2 and L3 in Fig. 8).

Due to the heterogeneous microstructure in additively manufactured 17-4PH steel and the oxygen content of the feedstock powder, the precipitation is expected to be different from its wrought counterpart. After aging at 482°C for 1 h, nanoscale Cu-rich clusters will act as the main strengthening phase in 17-4PH steels. The relationship between these precipitates and their coherency with the matrix needs to be clearly elucidated to understand the strengthening mechanism in AM 17-4PH steels. Figures 9 and 10 show the TEM micrographs of aged AM 17-4PH alloys after different homogenization conditions and subsequent aging at standard condition (482°C for 1 hour). For 0.5 hour homogenized alloy with standard aging, the width of martensite lath was refined (~300 nm) in comparison with the as-built alloy because of recrystallization that occurred during the short homogenization time. The dark particles seen in Figs. 9b and 9c correspond to the oxides that precipitated during homogenization. Besides, several pre-existing Cu precipitates (ε-Cu indicated with red arrows) from the as-built alloy were still remaining in the matrix, with an average size of 30 nm, incoherent with matrix as shown in Fig. 5f. These incoherent ε-Cu precipitates will allow the dislocations to bypass and hence, deteriorate the strength.

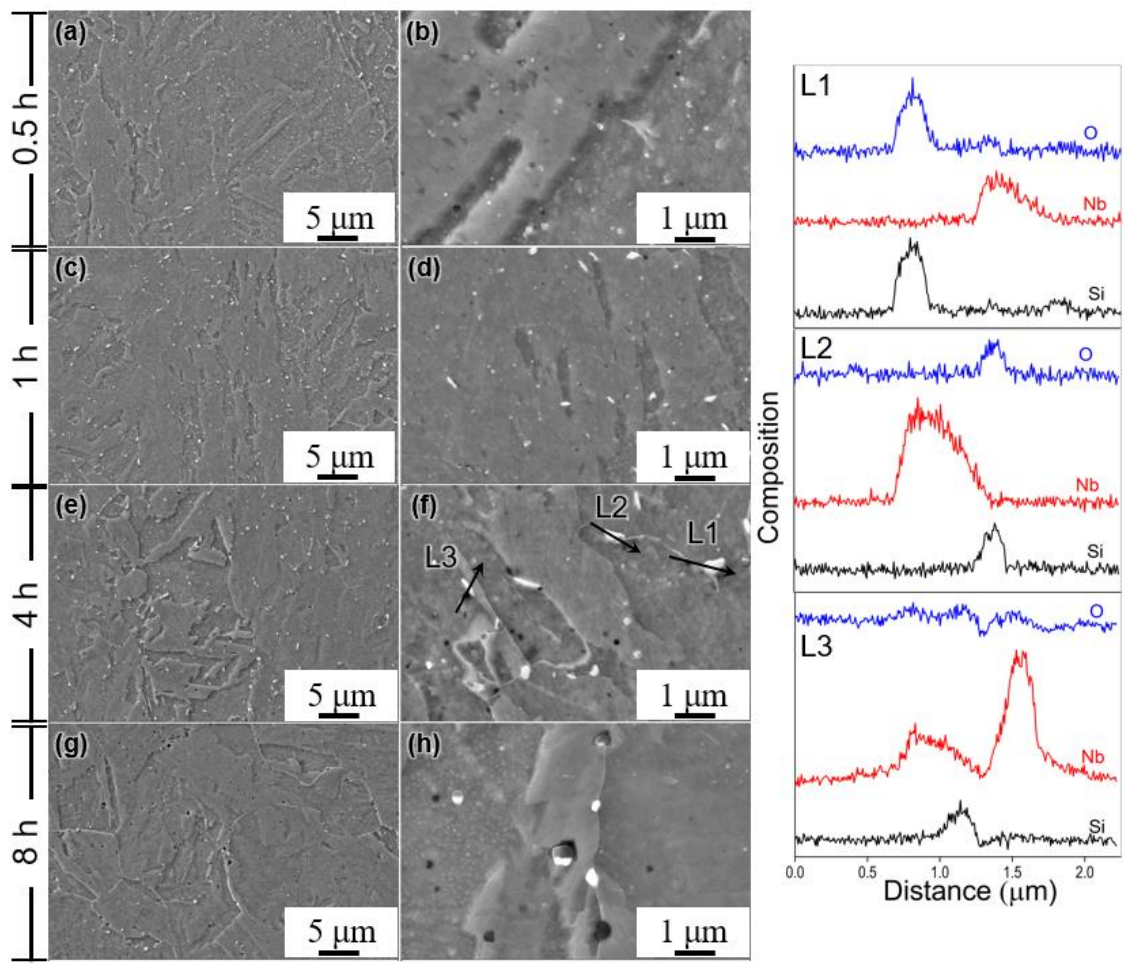

Fig. 8. SEM micrographs at different magnifications for 17-4PH steels after homogenizing for : (a, b) 0.5, (c, d) 1, (e, f) 4, (g, h) 8 hours. EDS composition profiles along L1, L2 and L3 in (f) are shown in the right column.

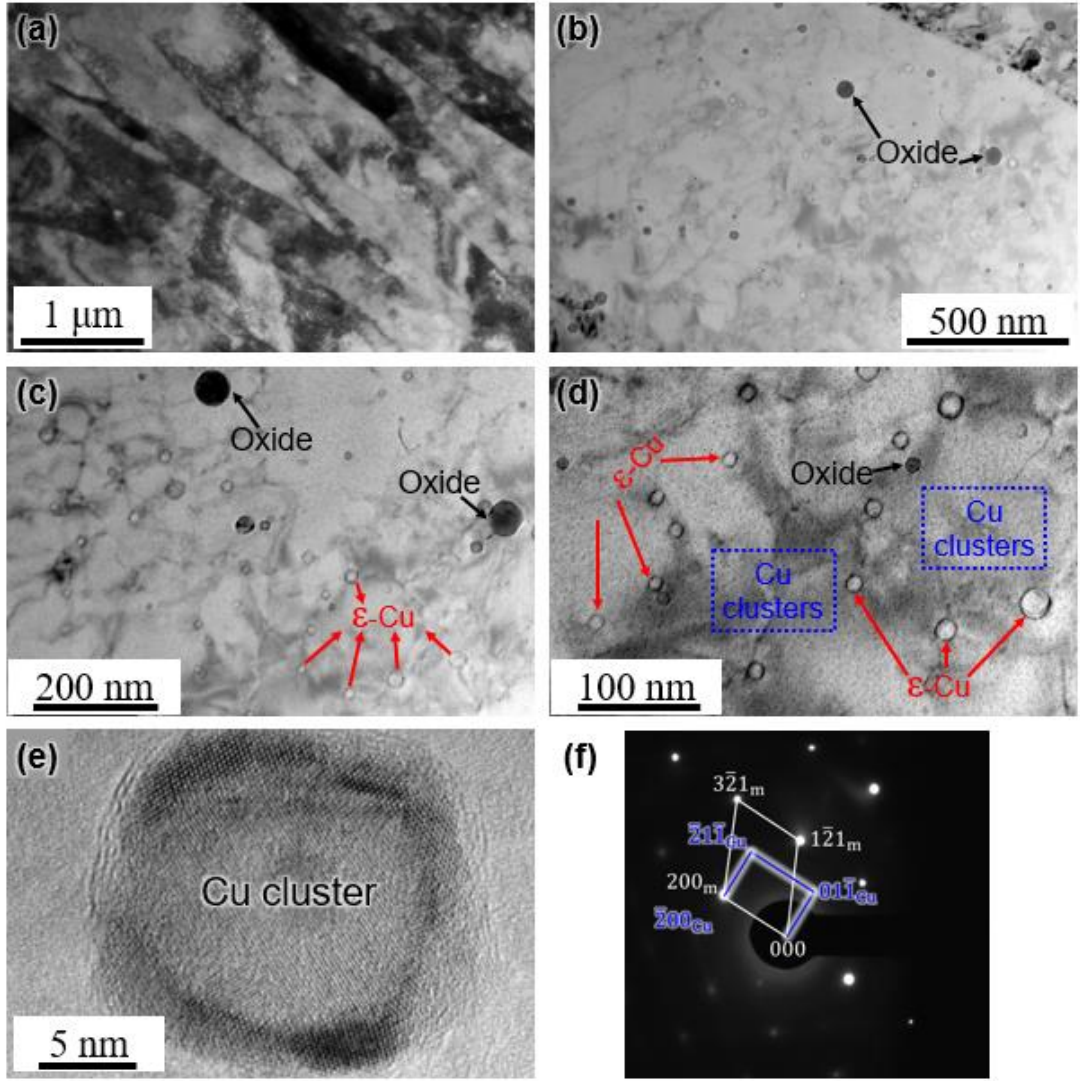

Fig. 9. TEM analysis for 0.5 hour homogenized 17-4PH alloy followed by standard aging: (a) morphology of lath martensite, secondary precipitates at (b) low and (c) high magnification, (d) high magnification micrograph of (c), (e) high resolution micrograph of Cu-rich cluster and (f) the corresponding selected area electron diffraction pattern of (e)

On the other hand, numerous small dark precipitates were densely dispersed in the matrix, as denoted by blue rectangles (Fig. 9d). Their size was only several nanometers. These particles are Cu-rich clusters which are the typical strengthening phases in 17-4PH steel during aging [3,11,46]. From the high-resolution TEM micrograph and SAED pattern, it is confirmed that the Cu-rich clusters are coherent with martensite matrix with the orientation relationship of $(200)_{CRPs} <011>_{CRPs}//(200)_m <012>_m$ (Figs. 9e and 9f). The Cu-rich clusters that are coherent with the matrix strengthens the material by shearing mechanism [1,46,48].

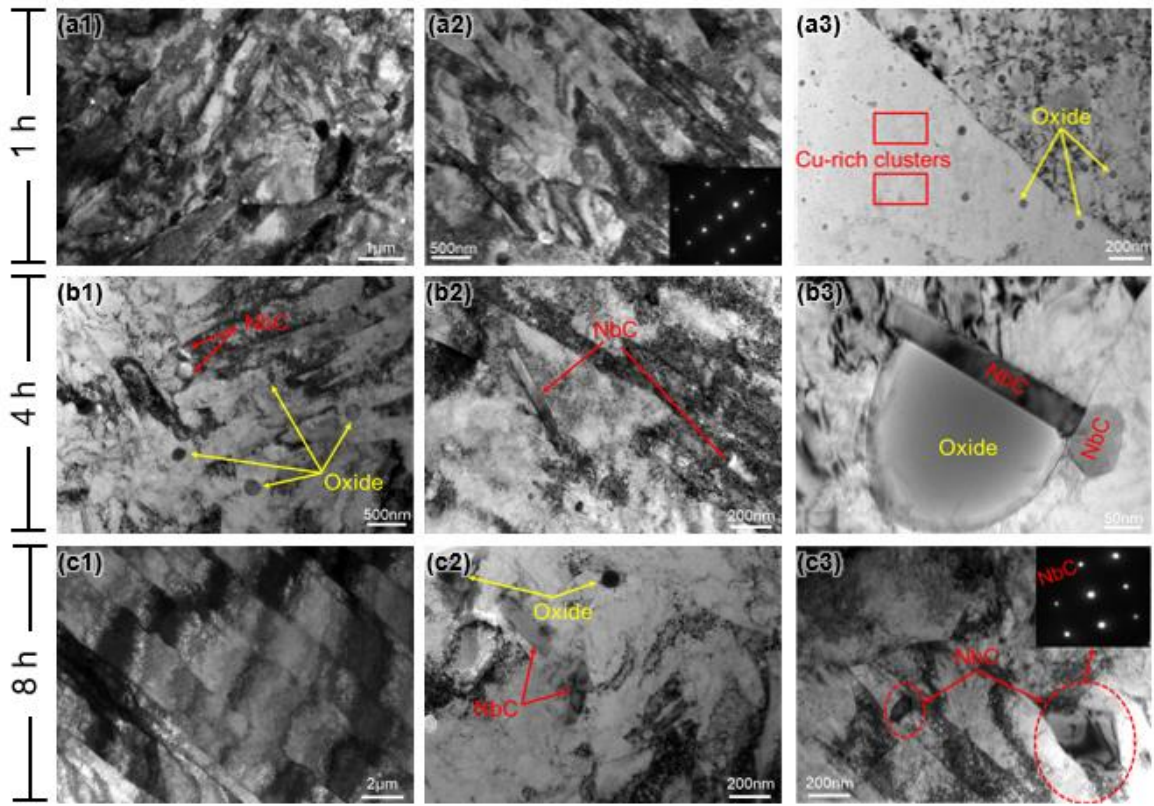

Figure 10. TEM micrographs for AM 17-4PH steels homogenized at different conditions and subsequently aged at the standard condition: (a) 1 hour homogenized showing martensite matrix (inset image showing the SAD pattern for martensite) with Cu-rich clusters and oxides, (b) 4 hours homogenized showing the presence of co-precipitated NbC and oxides and (c) 8 hours homogenized showing martensite matrix with coarsened NbC and oxide particles (inset image showing SAD pattern for NbC). 1-3 represent different areas of the sample.

Figure 10 a1-a3 illustrates the TEM micrographs from different zones in 1 hour homogenized alloy after standard aging at different magnifications. The martensite lath width reduced to 200 nm because of further recrystallization. The dislocation density in martensite decreased, which were observed from the SAED pattern of BCC structure (Fig. 10a2). Three types of precipitates namely, NbC, oxides and Cu-rich clusters, were observed in the matrix, as shown in Figs. 10a1 and 10a3. The ε-Cu precipitates disappeared, implying that 1 hour of homogenization was sufficient for their complete dissolution which explains why the hardness decreased rapidly during the initial homogenization without aging, even though NbC and oxides precipitated (Fig. 6). As the homogenization time increases from 1 to 4 hours, NbC and oxide particles formed along the grain boundaries thus, retarding the coarsening of PAG due to Zener pinning effect [29,60-63]. However, after 4 hours of homogenization followed by aging, NbC and oxides began to coarsen. NbC precipitates with needle-like and polygonal morphologies were observed in addition to co-precipitation with the oxide particles (Figs 10(b3)). Co-precipitation would result in crack formation during deformation due to huge difference in shear modulus between these two precipitates. After 8 hours of homogenization and subsequent aging, martensite and precipitates coarsened significantly. The size of NbC was greater than 200 nm, and the diffraction pattern of pure FCC structure Fig. 10(c3) that corresponds to that particle was observed.

### *3.3 Mechanical properties of additively manufactured 17-4PH steels*

Mechanical properties of AM 17-4PH alloys for four different representative periods of homogenization (0.5, 1, 4, and 8 h) followed by the standard aging heat treatment as well as the as-built alloy were determined from the engineering strain-stress curves obtained from tensile tests. The tensile properties are listed in Table 2 with their corresponding plots in Fig. 11. It is to be noted that a standard heat treatment was applied for the wrought 17-4PH steel with homogenization at 1038°C for 0.5 h and aging at 482°C for 1 h. The ultimate tensile strength (UTS: $\sigma_b$) can represent the strength and the static toughness ($U_{OT}$) can denote the toughness of the sample [56-58]. The static toughness was calculated from the integrated area of engineering strain-stress curve with the following equation [57]:

$$U_{OT} = \int_0^{\varepsilon_t} \sigma d\varepsilon \qquad (1)$$

where $U_{OT}$ is the static toughness, $\sigma$ is the real-time stress, $\varepsilon_t$ is the total strain after the fracture, $\varepsilon$ is the real-time strain. From Table 2, it is evident that the as-built sample has the lowest strength (UTS=922 MPa and yield strength (YS)=784 MPa) while its elongation was the highest (16.7%). This resulted in a high static toughness ($U_{OT}$) of 139.88 MJ·m$^{-3}$. After a short homogenization and standard aging, the strength increased (UTS: 1231MPa, YS: 1130 MPa), but the elongation (6%) and static toughness (69.39 MJ·m$^{-3}$) reduced considerably. After 4 hours of homogenization, the strength was highest with UTS of 1431 MPa and YS of 1309 MPa, while the elongation was 7.2%. However, there was a drop in the toughness (97.49 MJ·m$^{-3}$). With 1 hour of homogenization, the best trade-off of strength and elongation, with UTS of 1399 MPa and YS of 1280 MPa was obtained. The elongation and static toughness were found to be 10.5% and 140.50 MJ·m$^{-3}$, respectively. The tensile strength and ductility exceeded those of the wrought 17-4PH steel with standard heat treatment of homogenization at 1038°C for 0.5 hours followed aging at 482°C for 1 hour (UTS: 1379 MPa, YS: 1275 MPa, elongation: 9.0%). After prolonged homogenization for 8 hours, the strength reduced significantly whereas the ductility and static toughness improved slightly in comparison with the 4 hours homogenized alloy as shown in Fig. 9b.

Table 2. Tensile properties of AM 17-4PH steels with different homogenization conditions

| Sample | $\sigma_b$, MPa | $\sigma_s$, MPa | $\varepsilon_t$, % | $U_{OT}$, MJ·m$^{-3}$ |
|---|---|---|---|---|
| As-built | 922 | 784 | 16.7 | 139.88 |
| 0.5h Homogenized+Aged | 1231 | 1130 | 6.0 | 69.39 |
| 1h Homogenized+Aged | 1399 | 1280 | 10.5 | 140.50 |
| 4h Homogenized+Aged | 1431 | 1309 | 7.2 | 97.49 |
| 8h Homogenized+Aged | 1293 | 1165 | 9.0 | 111.26 |
| Wrought [5] | 1379 | 1275 | 9.0 | — |

*Note: $\sigma_b$ - ultimate tensile strength, $\sigma_s$ - yield strength, $\varepsilon_t$ - total elongation, $U_{OT}$ - static toughness.*

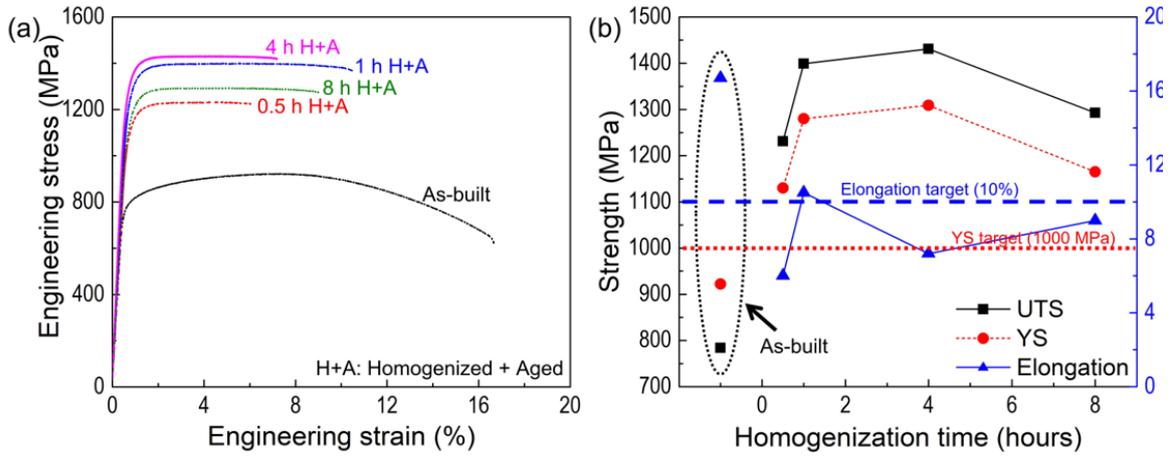

Fig. 11. Tensile properties of AM 17-4PH steels with different homogenization: (a) Engineering strain-

Figure 12 shows the fractographs for 17-4PH steels after different homogenization treatments followed by standard aging. For the as-built alloy, it had a fully ductile fracture surface, with dense and uniform dimples (Fig. 12 a1-a3). However, several shear bands and fracture walls along grain boundaries were observed (Fig. 12a1). This anisotropic fracture behavior is due to its heterogenous microstructure. At a high magnification, round shaped particles were found within the dimples, as indicated by blue dashed circles in Fig. 12a2. These particles are ε-Cu particles that are incoherent with the matrix (Fig. 5f) and hence, its strengthening effect is not comparable with the Cu-rich clusters that forms during aging. After 0.5 hour of homogenization followed by aging, the fracture surface transformed entirely to cleavage with typical river patterns (Fig. 12(b1)). Due to insufficient homogenization, the heterogenous microstructure was not completely removed. The pre-existing ε-Cu precipitates were remaining in the matrix, denoted with blue dashed circles in Fig. 12(b2). During deformation, the remaining ε-Cu precipitates and Cu-rich clusters formed during aging would lead to an easier initiation of microcracks, causing an early failure with low strength and ductility, as shown in Table 2 and Fig. 11a.

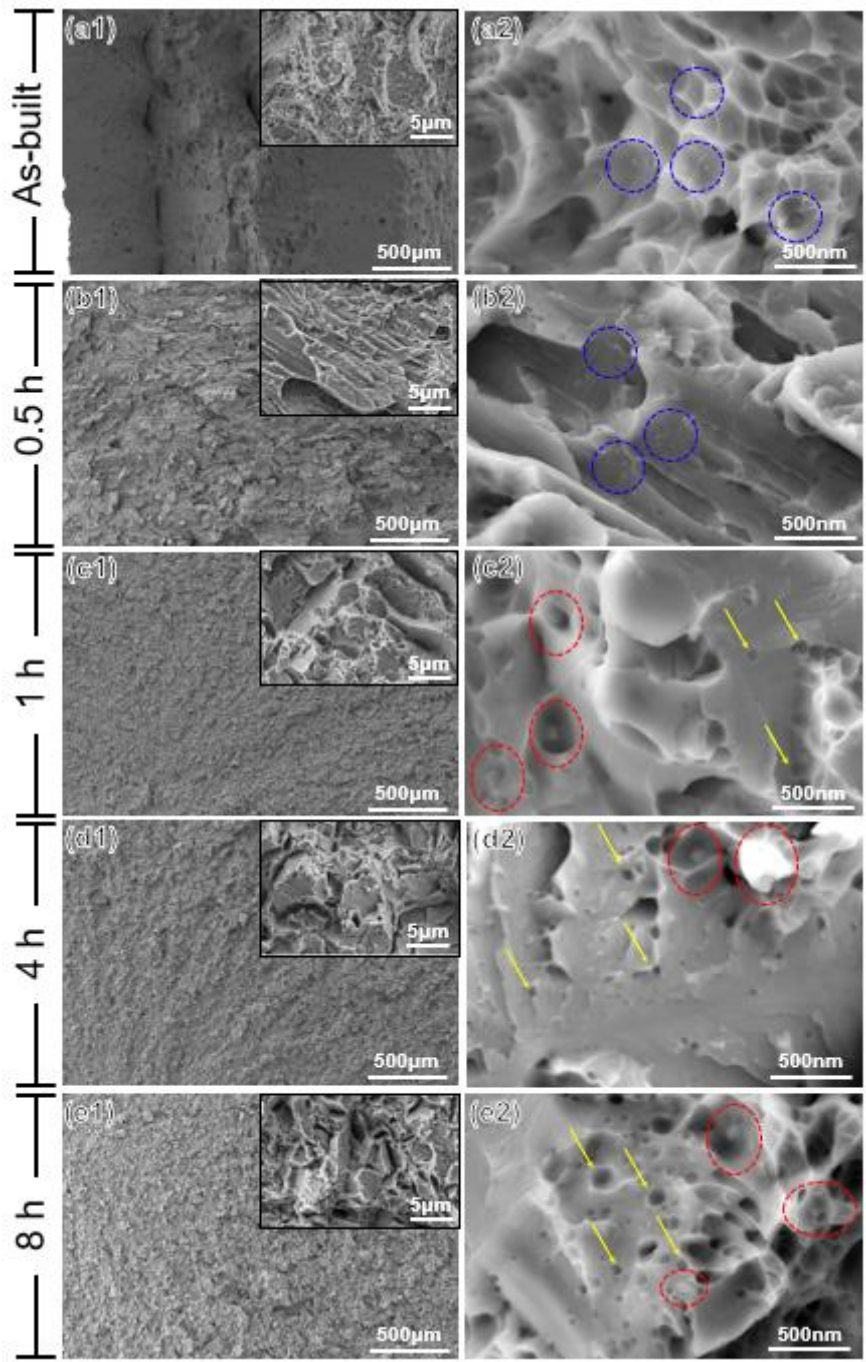

Figure 12. Fracture surfaces of 17-4PH steels in as-built condition and homogenized at 1050°C for different times followed by standard aging at different magnifications: (a) As-built, (b) 0.5, (c) 1, (d) 4 and (e) 8 hours.

When the homogenization time was longer than 1 h, the heterogenous microstructure was eliminated, leading to an isotropic fracture (Figs, 12(c1)-12(e1)). From the fracture surfaces at low magnification, it can be seen that the 1 hr homogenized alloy with standard aging had the finest fracture features, owing to its finely recrystallized grains. The insets of Figs. 12(c1)-12(e1) at a higher magnification shows that they had a mixed mode of fracture. The cleavage fracture is clearly seen between the grain/sub-grain boundaries, while each fracture surface of the grain/sub-grain has numerous dimples with several particles embedded within it (Figs. 12c2-12e2). Some particles reveal bright contrast and present within the dimples, as indicated by red dashed circles which correspond to the NbC. Several elliptical dark precipitates, as pointed out by yellow arrows denote the oxides. Comparing the fracture surfaces of three alloys homogenized from 1 to 8 hours followed by aging, it is clearly evident that the alloy homogenized for 4 hours followed by standard aging had the shallowest dimples, which results in the lowest ductility and highest strength (Fig. 11b). The dimples derived from NbC and oxides became bigger and were mixed together in the 8 hours homogenized alloy, because of the significantly coarsened NbC and oxide particles. It impaired the strength and ductility simultaneously as it can be seen from Table 2. In contrast, the 1 hour homogenized alloy with standard aging had a fracture surface with a dispersed distribution of precipitates. This is beneficial to both strength and ductility, which provided an excellent combination of mechanical properties. Hence, homogenization at 1050°C for 1 hour followed by aging at 482°C for 1 hour is found to display improved tensile properties in 17-4PH steels manufactured using LPBF technique. This necessitates the need for an optimized homogenization heat treatment for modifying the precipitation behavior during subsequent aging for AM 17-4PH steels to achieve enhanced strength and ductility.

## 5. Conclusions

- The post-heat treatment for 17-4PH steel fabricated by LPBF method has been designed successfully with improved mechanical properties in comparison with the wrought alloys.
- The oxygen content in powder feedstock can promote formation of oxide particles during printing while the cyclic heating and cooling generates ε-Cu precipitates. These precipitates are incoherent with the matrix in the as-built alloy thus, facilitating microcrack initiation during deformation.
- ε-Cu particles formed during printing can be dissolved completely after 1 hour of homogenization at 1050°C. 1 h homogenized alloy had the finest PAG and martensite width. Further aging at 482°C for 1 hour promotes the formation of Cu-rich clusters that are coherent with the matrix leading to strengthening. The UTS (1399 MPa) YS (1280 MPa) and ductility (10.5%) exceeds the corresponding tensile properties of wrought 17-4PH steels.
- NbC, nano-oxides and Cu-rich clusters play a critical role in strengthening the additively manufactured 17-4PH steels. Hence, this proves that altering the precipitation behavior using post-heat treatment is a promising way to improve the mechanical properties of additively manufactured alloys.

**CRediT authorship contribution statement**

**Kun Li**: Investigation, Methodology, Writing-original draft, Writing-review & editing; **Soumya Sridar:** Investigation, Writing-review & editing; **Susheng Tan**: Investigation, Methodology; Writing-review; **Wei Xiong**: Investigation, Methodology, Supervision, Writing-review & editing, Funding acquisition. The work done by **Kun Li** was accomplished entirely at the University of Pittsburgh as a postdoctoral associate.

**Declaration of Competing Interest**

The authors declare that they have no known competing financial interests or personal relationships that could have appeared to influence the work reported in this paper.

**Acknowledgements**

Xiong Group (K.L., S.S. and W.X.) gratefully acknowledges the support from the Department of Energy, Nuclear Energy University Program (NEUP Project 18-15251). Any opinions, findings and conclusions or recommendations expressed in this material are those of the author(s) and do not necessarily reflect those of the Department of Energy. W.X. acknowledge the support of the software and databases from Thermo-Calc Software Company through the ASM Materials Genome Toolkit Award.

**Notes for manuscript releasing as a pre-print**

This manuscript was submitted to a peer-reviewed journal in March, 2021. The original work was performed at the University of Pittsburgh by Kun Li in 2020 under the supervision of Wei Xiong. Susheng Tan supported the TEM study and manuscript preparation. The manuscript received rejection due to lengthy discussion, and the first author Kun Li has moved to Chongqing University, China, without being able to perform the further revision. Later, according to the authors' agreement, Soumya Sridar continued working on the manuscript revision. S.S. cut down the manuscript size significantly (almost half), re-performed modeling, and revisited data analysis. By Nov. 2021, S.S., S.T., and W.X. think this manuscript can be resubmitted. However, K.L. still thinks more revisions are required. Since the authors cannot reach an agreement for resubmission, a pre-print like this can help disseminate results.